\documentclass[aps,prl,reprint,superscriptaddress,amsmath,amssymb]{revtex4-2}
\usepackage{graphicx}
\usepackage{longtable}
\usepackage{xcolor}
\usepackage[bookmarks,bookmarksnumbered,colorlinks,allcolors=blue]{hyperref}
\usepackage{bookmark}
\usepackage{ifthen}
\usepackage{ulem}
\usepackage{soul}

\newcommand{\chenlu}[1]{\textcolor{black}{#1}}

\newcommand{\bra}[1]{\langle{#1}|}
\newcommand{\ket}[1]{|{#1}\rangle}

\begin{document}
\title{Converting qubit relaxation into erasures with a single fluxonium}

\author{Chenlu Liu}
\thanks{These authors contributed equally to this work}
\affiliation{Quantum Science Center of Guangdong-Hong Kong-Macao Greater Bay Area, Shenzhen, China}

\author{Yulong Li}
\thanks{These authors contributed equally to this work}
\affiliation{Beijing Key Laboratory of Fault-Tolerant Quantum Computing, Beijing Academy of Quantum Information Sciences, Beijing 100193, China}

\author{Jiahui Wang}
\author{Quan Guan}
\author{Lijing Jin}
\author{Lu Ma}
\author{Ruizi Hu}
\author{Tenghui Wang}
\author{Xing Zhu}
\affiliation{Quantum Science Center of Guangdong-Hong Kong-Macao Greater Bay Area, Shenzhen, China}

\author{Hai-Feng Yu}
\affiliation{Beijing Key Laboratory of Fault-Tolerant Quantum Computing, Beijing Academy of Quantum Information Sciences, Beijing 100193, China}
\affiliation{Hefei National Laboratory, Hefei 230088, China}

\author{Chunqing Deng}
\email{dengchunqing@quantumsc.cn}
\affiliation{Quantum Science Center of Guangdong-Hong Kong-Macao Greater Bay Area, Shenzhen, China}

\author{Xizheng Ma}
\email{maxizheng@quantumsc.cn}
\affiliation{Quantum Science Center of Guangdong-Hong Kong-Macao Greater Bay Area, Shenzhen, China}

\begin{abstract}
\textbf{
Qubits that experience predominantly erasure errors offer distinct advantages for fault-tolerant operation. Indeed, dual-rail encoded erasure qubits in superconducting cavities and transmons have demonstrated high-fidelity operations by converting physical-qubit relaxation into logical-qubit erasures, but this comes at the cost of increased hardware overhead and circuit complexity. Here, we address these limitations by realizing erasure conversion in a single fluxonium operated at zero flux, where the logical state is encoded in its $\ket{0}$-$\ket{2}$ subspace. A single, carefully engineered resonator provides both mid-circuit erasure detection and end-of-line (EOL) logical measurement.
Post-selection on non-erasure outcomes results in more than four-fold increase of the logical lifetime, from $193~\mu$s to $869~\mu$s. 
Finally, we characterize measurement-induced logical dephasing as a function of measurement power and frequency, and infer that each erasure check contributes a negligible error of $7.2\times 10^{-5}$. These results establish integer-fluxonium as a promising, resource-efficient platform for erasure-based error mitigation, without requiring additional hardware.}

\end{abstract}

\maketitle

Recent years have seen remarkable progress toward realizing fault-tolerant quantum computation. Multiple platforms, such as superconducting circuits~\cite{GoogleQuantumAI2021, GoogleQuantumAI2023,GoogleQuantumAI2025, zuchongzhi}, trapped ions~\cite{Moses2023, RyanAnderson2024, Liu2025}, and neutral atoms~\cite{Evered2023,Bluvstein2024,Bluvstein2026}, are now routinely assembling and controlling tens to hundreds of physical qubits at or beyond the error threshold required for quantum error correction. In particular, superconducting systems have surpassed the break-even point using the surface code, demonstrating logical qubits that outperform their constituent physical qubits~\cite{GoogleQuantumAI2025, Tan2025}. As these systems mature, gains in physical qubit performance have become increasingly incremental. Consequently, achieving substantial reductions in logical error rates
depends on a significant increase in system size. This challenge has propelled the search for hardware-efficient error-correction strategies, including more efficient codes~\cite{Cohen2022,Bravyi2024}, qubits exploiting structured or biased noise processes~\cite{Lescanne2020, Grimm2020, Putterman2025, tuckett2019, Darmawan2021}, and encoding redundancy directly into physical qubits using additional energy levels~\cite{Campbell2014,Li2024_AQEC,Goss2024_qutrit}.

Among these approaches, erasure qubits~\cite{Wu2022,Scholl2023,Ma2023,Teoh2023,Kubica2023} have recently attracted considerable attention, offering a particularly hardware-efficient pathway to fault tolerance by converting dominant physical errors into erasures. These are leakage errors whose occurrence can be detected in both time and space without disturbing the quantum information encoded in the computational subspace. 
Knowing the precise error locations greatly simplifies decoding and makes erasure errors easier to correct. In surface code, for example, erasure errors can be corrected at a substantially lower error threshold and allow a code of fixed distance to correct roughly twice as many errors~\cite{Wu2022,Kubica2023}, thereby reducing the required hardware overhead.

In superconducting circuits, erasure qubits have been realized using dual-rail encodings of pairs of transmons~\cite{Levine2024,Huang2025} or superconducting cavities~\cite{Chou2024,deGraaf2025,mehta2025}. Logical states are defined in the single-excitation manifold of each qubit pair, so that the decay of any physical qubit is converted into a detectable leakage event outside the logical subspace. Mid-circuit detection~\cite{Koottandavida2024, deGraaf2025,Levine2024} of such events enables high-fidelity single-qubit and entangling gates~\cite{Huang2025,mehta2025} that preserve the biased error structure of the system, paving the way for integration of superconducting erasure qubits into higher-level error-correction codes. However, current implementations require multiple physical qubits to encode a single logical qubit, along with additional qubits for mid-circuit or logical detection. This introduces considerable hardware overhead and circuit complexity, offsetting the efficiency gained from erasure conversion. Although proposals~\cite{Kubica2023} exist for realizing erasure qubits within a single physical qubit circuit, experimental investigations have so far remained limited.

\begin{figure}[h!]
    \centering
    \includegraphics{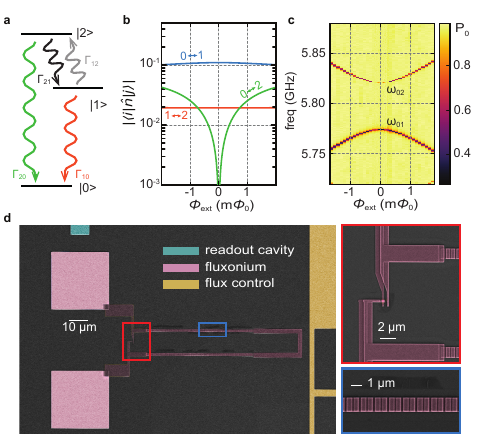}
    \caption{\textbf{Erasure conversion in a single fluxonium.} 
    \textbf{a,} The logical qubit is encoded in the $\{\ket{0},\ket{2}\}$ subspace of an integer fluxonium. The lowest three energy eigenstates are connected by spontaneous transitions $\ket{i}\rightarrow \ket{j}$ at rate $\Gamma_{ij}$. To facilitate erasure conversion, the direct transition rate $\Gamma_{20}$ should be ideally zero, such that relaxations from $\ket{2}$ always proceed through $\ket{1}$ as an intermediate state. This can be achieved by operating at the zero-flux position.
    \textbf{b,} The vanishing charge-matrix element $|\bra{0}\hat{n}\ket{2}|$ at $\Phi_\text{ext} = 0$ testifies to a symmetry-induced suppression of the direct transition, which establishes the basis for erasure conversion.
    \textbf{c,} The qubit spectrum likewise reflects this symmetry-induced suppression: the spectral feature corresponding to the $\ket{0}\leftrightarrow\ket{2}$ transition vanishes at $\Phi_\text{ext} = 0$.
    \textbf{d,} False-colored scanning electron micrograph of the device. The insets provide closer views of the overlay junctions (blue box), which form the junction loop, and the Manhattan junction (red box), which provides the Josephson energy, at locations indicated by the boxes of the corresponding colors.
    }
    \label{fig:Fig1}
\end{figure}

In this work, we demonstrate that erasure conversion can be realized within a single physical qubit circuit. Specifically, as shown in \autoref{fig:Fig1}\textbf{a}, we encode a logical qubit in the $\{\ket{0}$,$\ket{2}\}$ subspace of a fluxonium operated at zero magnetic flux~\cite{Ardati2024,Mencia2024}, thereby converting the intermediate $\ket{1}$ state into a detectable erasure channel. 
To facilitate erasure conversion, any decay connecting the logical states $\ket{0}$ and $\ket{2}$ must pass predominantly through the erasure state. In other words, the direct decay rate between the two logical states, $\Gamma_{20}$, should ideally vanish. For a fluxonium qubit described by the Hamiltonian,
\begin{equation}
    H = 4E_C \hat{n}^2 - E_J \cos{\hat{\varphi}} + \frac{E_L}{2}\left(\hat{\varphi} + 2\pi \frac{\Phi_\text{ext}}{\Phi_0}\right)^2,
\end{equation}
where $E_C$, $E_L$, and $E_J$ are the charging, inductive, and Josephson energies, and $\hat{n}$ and $\hat{\varphi}$ are the Cooper-pair number and phase operators, this condition can be satisfied by operating at either zero flux, $\Phi_\text{ext} = 0$, or at the half flux-quantum, $\Phi_\text{ext} = \Phi_0/2$. As an example, \autoref{fig:Fig1}\textbf{b} shows the vanishing charge matrix element $|\bra{0}\hat{n}\ket{2}|$ at $\Phi_\text{ext} = 0$.
Because this matrix element governs the direct transition rate between qubit states according to Fermi's Golden Rule, $\Gamma_{ij} \propto |\bra{i}\hat{n}\ket{j}|^2$, the symmetry-induced suppression of $|\bra{0}\hat{n}\ket{2}|$ therefore forbids direct $\ket{0}\leftrightarrow\ket{2}$ transitions and ensures that relaxations from $\ket{2}$ always proceed through the intermediate state $\ket{1}$. 

Here, we work with an integer fluxonium~\cite{Ardati2024,Mencia2024} operated at $\Phi_\text{ext} = 0$. Spectroscopy near the operating point (\autoref{fig:Fig1}\textbf{c}) enables the extraction of device parameters $E_c/\hbar = 2\pi \times 1.72~\text{GHz}$, $E_J/\hbar = 2\pi \times 7.07~\text{GHz}$, and $E_L/\hbar = 2\pi \times 0.32~\text{GHz}$. In contrast to a standard~\cite{Bao2022,Ding2023} fluxonium operated at $\Phi_\text{ext} = \Phi_0/2$, which typically has qubit transition frequency $\omega_{01}$ in the tens to hundreds MHz range and $\omega_{12}$ in the GHz, our integer fluxonium (\autoref{fig:Fig1}\textbf{d}) exhibits the opposite spacing, with $\omega_{01} = 2\pi \times 5.77~\text{GHz}$ much larger than $\omega_{12} = 2\pi \times 48.0~\text{MHz}$. This energy landscape provides several advantages:
The small $\omega_{12}$ reduces the susceptibility of $\ket{2}$ to dielectric loss~\cite{sun2023}, enabling a substantially longer lifetime and a lower physical error rate.
Although the larger $\omega_{01}$ would reduce the dielectric-loss limit of the $\ket{1}$ lifetime, experiments~\cite{Ardati2024,Mencia2024} show it could remain above $100~\mu\text{s}$, sufficiently long for erasure detections. 
Moreover, the large $\omega_{01}$ ensures that the qubit predominantly thermalizes to $\ket{0}$ at dilution refrigerator base temperatures, facilitating straightforward initialization and reset.

\begin{figure}[h!]
    \centering
    \includegraphics{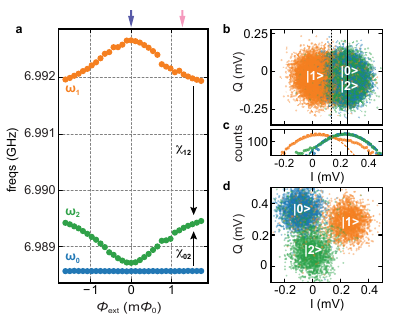}
    \caption{\textbf{Erasure detection and end-of-line (EOL) measurement.}
    \textbf{a,} The qubit dressed resonator frequency $\omega_i$ is measured as a function of $\Phi_\text{ext}$ for the lowest three energy eigenstates of the qubit. We exploit the resulting flux dependence of the dispersive shifts $\chi_{ij} = \omega_i - \omega_j$ to enable erasure detection at $\Phi_\text{ext} = 0$ (purple arrow) and EOL measurement at \chenlu{$\Phi_\text{ext} = 1.3~m\Phi_0$} (pink arrow). 
    \textbf{b, c,} With an average intra-cavity photon number of $n = 2.3$ and an integration time of $t_\text{meas} = 1.6~\mu$s, the erasure detection cannot distinguish $\ket{0}$ from $\ket{2}$ in the IQ-plane (\textbf{b}), but resolves $\ket{1}$ clearly. Rather than using the midpoint (dotted vertical line) between the state distributions, we intentionally bias the assignement threshold to the right (solid vertical line) to reduce the probability of misidentifying $\ket{1}$ as a computational state (false-negative), at the cost of increasing the chance of erroneously labeling a computational state as an erasure event (false-positive). With this choice, the histogram (\textbf{c}) reveals a false-negative rate of $4.9\%$ and a false-positive rate of $56.7\%$. 
    \textbf{d,} At \chenlu{$\Phi_\text{ext} = 1.3~m\Phi_0$}, with $n = 5.6$ and the same $t_\text{meas} = 1.6~\mu$s, the EOL measurement resolves all three states clearly.
    }
    \label{fig:Fig2}
\end{figure}

Operating an erasure qubit requires two distinct types of measurements: mid-circuit erasure detection and end-of-line (EOL) logical readout. The former must distinguish the erasure state $\ket{1}$ from the computational subspace $\{\ket{0}, \ket{2}\}$ without disturbing the encoded information, while the latter aims to resolve all three states.
We implement both measurements using a single resonator dispersively coupled to the qubit. In this regime, each qubit state $\ket{i}$ shifts the resonator frequency to a distinct value $\omega_i$, thereby enabling state discrimination. A large net dispersive shift, $\chi_{ij} = \omega_i - \omega_j$, allows the states $\ket{i}$ and $\ket{j}$ to be distinguished during measurement, but also reveals information about the subspace, resulting in additional dephasing~\cite{Gambetta2006} of the encoded quantum state. Consequently, for logical readout, all $\chi_{ij}$ must be large enough to resolve $\ket{0}$, $\ket{1}$, and $\ket{2}$. In contrast, for an erasure measurement, $\chi_{01}$ and $\chi_{12}$ should be sizable to identify the erasure state, while $\chi_{02}$ should be ideally zero to avoid perturbing the computational subspace. 
To perform both measurements using a single resonator, we exploit the flux dependence of the resonator's dispersive shift near $\Phi_\text{ext} = 0$ (\autoref{fig:Fig2}\textbf{a}).

At the operation point $\Phi_\text{ext} = 0$, our device is carefully designed for erasure conversion, with $\chi_{02} = -2\pi \times 147~\text{kHz}$, more than an order of magnitude smaller than \chenlu{$\chi_{01} = - 2\pi \times 4.096~\text{MHz}$}. This large contrast enables clear identification of the erasure state $\ket{1}$ without revealing information about $\{\ket{0}, \ket{2}\}$. Indeed, as shown in \autoref{fig:Fig2}\textbf{b, c}, with an integration time of $t_\text{meas} = 1.6~\mu$s, the readout signal of $\ket{0}$ largely overlaps with that of $\ket{2}$, whereas $\ket{1}$ is clearly distinguishable.
The vertical lines indicate the selection thresholds, with measurement outcomes lying to their right labeled as erasure events. Limited by a moderate readout efficiency~\cite{Bultink2018} of $29.8\%$ and a small photon number, $n = 2.3$, chosen to preserve the quantum nondemolition nature of the measurement (see Supplementary Material), the maximal fidelity to distinguish $\ket{1}$ from the computational manifold $\{\ket{0},\ket{2}\}$ is $86.9\%$ when the threshold is set to the midpoint of the corresponding distributions (dotted vertical line). However, to minimize the chance of missing an actual erasure event, we intentionally bias the threshold toward the right (solid vertical line) -- preferring to misidentify computational states as erasures rather than to overlook genuine error events. With this choice, each erasure-conversion measurement yields a false-negative rate of $4.9\%$ (missing an erasure event) but a false-positive rate of $56.7\%$ (computational state mislabeled as erasure).

For EOL logical readout, by contrast, we pulse the qubit away from $\Phi_\text{ext} = 0$. Specifically, we perform the measurement at \chenlu{$\Phi_\text{ext} = 1.3~m\Phi_0$}, where the dispersive shifts \chenlu{$\chi_{02} = - 2\pi \times 760~\text{kHz}$} and \chenlu{$\chi_{01} = - 2\pi \times 3.503~\text{MHz}$} are both sufficiently large to enable simultaneous discrimination of all three states (\autoref{fig:Fig2}\textbf{d}) with a fidelity of $86.1\%$. While further increasing the flux displacement could improve the separation between these states, it would also introduce additional state-transition errors due to increased decoherence. 

\begin{figure*}[t]
    \centering
    \includegraphics{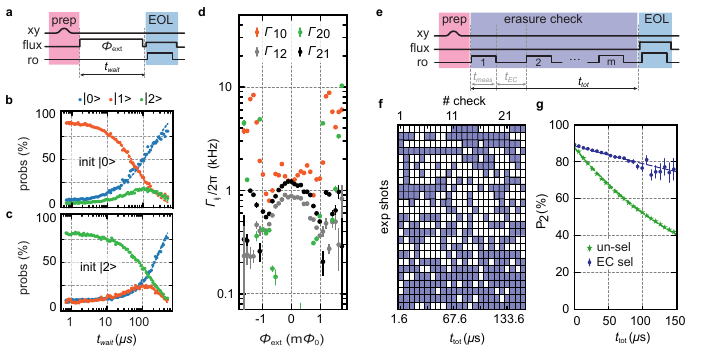}
    \caption{\textbf{Converting decay errors into erasures. a,} The dynamics of the qubit relaxation are measured by idling the qubit at a fixed $\Phi_\text{ext}$ for a variable wait time $t_\text{wait}$ after initialisation. The time-dependent populations of the lowest three energy states are then acquired via an EOL logical measurement. 
    \textbf{b, c,} At $\Phi_\text{ext} = 0$, for example, the spontaneous transition rates $\Gamma_{ij}$ can be extracted by jointly fitting the measured dynamics for the qubit initially prepared in $\ket{1}$ (\textbf{b}) and $\ket{2}$ (\textbf{c}). 
    \textbf{d}, Repeating this procedure for different $\Phi_\text{ext}$ maps out the transition-rate landscape of $\Gamma_{ij}$. Importantly, close to $\Phi_\text{ext} = 0$, the direct transition rate $\Gamma_{20}$ between $\ket{2}$ and $\ket{0}$ is suppressed by orders of magnitude compared to all other rates. This ensures that logical decays within the computational space predominantly occur via $\ket{1}$ as an intermediate state, thereby enabling the conversion of decay events into detectable erasures.
    \textbf{e,} To characterise erasure conversion, a qubit initialized in $\ket{2}$ is subjected to a series of $m$ erasure detections, each lasting $t_\text{meas} = 1.6~\mu$s and separated by a variable interval $t_\text{EC}$, resulting in a total delay of $t_\text{tot}$, before an EOL measurement is performed.
    \textbf{f,} A sample of erasure-detection outcomes for $t_\text{EC} = 5~\mu$s and $t_\text{tot} = 153.4~\mu$s is shown, where each row corresponds to a single experimental shot, and a purple-colored block on the $i$-th column indicates the detection of an erasure event during the $i$-th detection.
    \textbf{g,} Compared to ignoring the erasure information (green), post-selecting on traces where no erasure event is detected (purple) extends the characteristic logical lifetime by more than a factor of four, from \chenlu{$193~\mu$s to $869~\mu$s}.
    }
    \label{fig:Fig3}
\end{figure*}

We proceed to characterize the relaxation rates among the lowest three energy eigenstates of the qubit and confirm the hierarchy required for erasure conversion. Initialized in either $\ket{1}$ or $\ket{2}$ state, the qubit is idled for a variable wait time $t_{\mathrm{wait}}$ at a fixed $\Phi_\text{ext}$. We then extract and record the time-dependent populations of all three states using an EOL measurement (\autoref{fig:Fig3}\textbf{a}). 
The dynamics of these states can be modeled by the set of classical master equations,
\begin{equation}
\begin{aligned}
dP_0/dt &= \Gamma_{10} P_1 + \Gamma_{20} P_2, \\
dP_1/dt &= -(\Gamma_{10} + \Gamma_{12}) P_1 + \Gamma_{21} P_2, \\
dP_2/dt &= -(\Gamma_{20} + \Gamma_{21}) P_2 + \Gamma_{12} P_1,
\end{aligned}
\label{eq:three_state_component}
\end{equation}
where $P_i$ is the instantaneous population in the qubit $\ket{i}$ state ($P_0 + P_1 + P_2 = 1$), and $\Gamma_{ij}$ describes the spontaneous transition rate for the process $\ket{i} \rightarrow \ket{j}$. In this equation, we neglect spontaneous excitations from $\ket{0}$, because the large energy gaps $\omega_{01}$ and $\omega_{02}$ make such events exceedingly rare for a qubit thermalized to the base temperature of a dilution refrigerator.
As an example, \autoref{fig:Fig3}\textbf{b, c} shows the measured relaxation (dots) at $\Phi_\text{ext} = 0$ when the qubit is initialized in $\ket{1}$ and $\ket{2}$, respectively. A joint fit to the master equations (solid lines) reveals the spontaneous transition rates. 

Repeating this procedure for different $\Phi_\text{ext}$, \autoref{fig:Fig3}\textbf{d} maps the landscape of the qubit's spontaneous transition rates. In general, the small transition frequency $\omega_{12} \ll \omega_{01}$ suppresses the dielectric loss in the $\{\ket{1},\ket{2}\}$ manifold, leading to a smaller emission rate $\Gamma_{21}$ compared to $\Gamma_{10}$. At the same time, the small $\omega_{12}$ also leads to a significant thermal occupation in the manifold $\{\ket{1},\ket{2}\}$, such that detailed balance enforces a comparable absorption rate, $\Gamma_{12} \lesssim \Gamma_{21}$. Importantly, while the direct $\ket{2} \rightarrow \ket{0}$ transition rate $\Gamma_{20}$ is comparable to the other rates when the qubit is biased away from the operating point, it becomes strongly suppressed near $\Phi_{\text{ext}} = 0$. 
Indeed, at $\Phi_{\text{ext}} = 0$, we find $\Gamma_{10} = 2\pi\times 1.22 ~\text{kHz}$, $\Gamma_{21} = 2\pi\times 1.21~\text{kHz}$, and $\Gamma_{12} = 2\pi\times 0.88~\text{kHz}$, while $\Gamma_{20}$ is more than an order of magnitude smaller and could not be reliably extracted from the joint fit. Nevertheless, we include all the extracted $\Gamma_{20}$ values in the figure to illustrate the strong suppression trend, even if their absolute numbers around $\Phi_\text{ext} = 0$ may not be quantitatively accurate. 

This rate hierarchy ensures that decay events within the computational subspace almost always pass through $\ket{1}$ as an intermediate step, which we can exploit to convert the decay events into erasures by monitoring $\ket{1}$.
\autoref{fig:Fig3}\textbf{e} outlines our measurement sequence. Initialized in $\ket{2}$, the qubit undergoes $m$ rounds of erasure detections, each lasting $t_\text{meas} = 1.6~\mu$s and separated by a variable interval $t_\text{EC}$, for a total duration $t_\text{tot} = m(t_\text{meas} + t_\text{EC})$. The sequence concludes with an EOL measurement to extract the qubit state.
\autoref{fig:Fig3}\textbf{f} shows the erasure-detection outcomes for $t_\text{EC} = 5~\mu$s and a total duration of \chenlu{$t_\text{tot} = 153.4~\mu$s}, with each row representing a single experimental shot. The first trace illustrates a near-ideal trajectory, in which the qubit stays in $\ket{2}$ until the $11$-th shot, where the onset of a decay event within the computational subspace is marked by the detection of $\ket{1}$ (purple blocks).
The qubit subsequently resides in $\ket{1}$ for an additional $5$ shots, before either relaxing to $\ket{0}$ or being thermally re-excited to $\ket{2}$. The large false-positive rate of our erasure detection means that, during each measurement, a qubit remaining in the computational subspace has a $56.7\%$ chance of being misidentified as being in the erasure state. This results in the apparent rapid switching between states on timescales that are short compared to the qubit's spontaneous transition times.

Nevertheless, for a given $t_\text{EC}$ and $t_\text{tot}$, we post-select based on the erasure detection results, discarding any runs in which the qubit was ever found in $\ket{1}$. \autoref{fig:Fig3}\textbf{g} then shows the probability of finding the qubit in $\ket{2}$ in the EOL measurement. 
Compared to the unselected data (green), post-selection (purple) clearly extends the effective lifetime of $\ket{2}$. In principle, an erasure event could go undetected if the qubit were to return to the computational space before the next erasure check. This occurs on a characteristic timescale $T_{\ket{1}} = 1/(\Gamma_{10} + \Gamma_{12})\approx 75.8~\mu$s. 
Consequently, a shorter $t_\text{EC}$ reduces the probability of missing an erasure and leads to an improved effective lifetime (see Supplementary Material). 
Although these trajectories do not follow a simple exponential decay, we nonetheless extract a characteristic logical lifetime $T_\text{1L}$ to quantify the improvement. 
For $t_\text{EC} = 5~\mu$s, we find that post-selecting on erasure events increases $T_\text{1L}$ by more than four-fold, from \chenlu{$193~\mu$s} to \chenlu{$869~\mu$s}. Ultimately, this lifetime is likely limited by a combination of missed erasure events, measurement-induced errors, such as false-negatives and non-QND errors, and direct transitions within the computational subspace.

\begin{figure}
    \centering
    \includegraphics{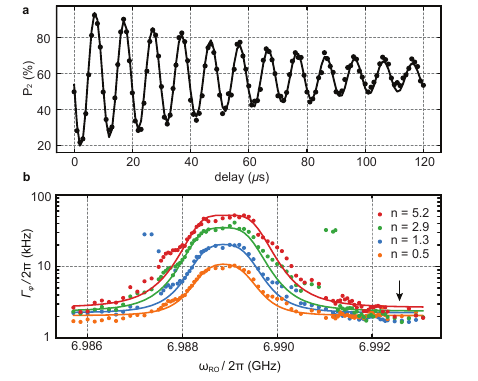}
    \caption{\textbf{Dephasing of the logical qubit}. 
    \textbf{a,} We measure the decoherence of the logical qubit using a Ramsey sequence, and extract a coherence time $T_\text{2L} = 70.4~\mu$s. 
    \textbf{b,} Applying continuous erasure detections during the Ramsey free-evolution period, we extract the dephasing rate $\Gamma_\varphi = 1/T_\text{2L} - 1/(2T_\text{1L})$ as a function of readout frequency $\omega_\text{RO}$ and intra-cavity photon number $n$. The measured data (dots) are in good agreement with the theory of measurement-induced dephasing (solid lines). For erasure detections used in \autoref{fig:Fig2} and \autoref{fig:Fig3}, with $n = 2.3$, $t_\text{meas} = 1.6~\mu$s, and $\omega_\text{RO} = 2\pi\times 6.993~$GHz (vertical arrow), we find a dephasing error of $\epsilon_\varphi \approx 7.2\times 10^{-5}$ per check.
    }
    \label{fig:Fig4}
\end{figure}

\autoref{fig:Fig4}\textbf{a} presents the measured coherence of logical states within the $\{\ket{0},\ket{2}\}$ subspace. The Ramsey sequence is performed by applying a logical $\pi/2$ pulse, implemented via a microwave-activated two-photon transition (see Supplementary Material), followed, after a variable delay, by a second logical $\pi/2$ pulse before the qubit is measured using the EOL measurement. 
In our device, the logical coherence time $T_\text{2L} = 70.4~\mu$s is unfortunately short compared to the characteristic time for an erasure error to occur, $T_\text{eras} = 1/\Gamma_{21} \approx 132 ~\mu$s. As a result, the detection and post-selection of erasure errors does not significantly improve $T_\text{2L}$.

Finally, we quantify the impact of repeated erasure detection on qubit dephasing by applying continuous erasure detections during the Ramsey free-evolution period. The dephasing rate is extracted as $\Gamma_\varphi = 1/T_{2\mathrm{L}} - 1/(2T_{1\mathrm{L}})$ and is shown in \autoref{fig:Fig4}\textbf{b} (dots) as a function of the readout power and frequency of the erasure measurement. These results are in good agreement with the theory of measurement-induced dephasing (solid lines, see Supplementary Material)~\cite{Gambetta2006}.  From the fit, we estimate that the erasure detection characterized in \autoref{fig:Fig2} and used in \autoref{fig:Fig3} introduces a dephasing error of $\epsilon_\varphi \approx 7.2 \times 10^{-5}$ per check (see Supplementary Material). This negligible amount of added error makes frequent monitoring of erasure state practical and establishes integer-fluxonium qubits as a viable platform for erasure-based error mitigation.

In summary, our experiment represents a first step toward an erasure-based error correction architecture using single integer-fluxonium qubits. Compared with dual-rail architectures, the main advantage of our approach is its minimal requirement for quantum resources. Using a single qubit coupled to a single resonator, we convert qubit decay errors into detectable erasures and extend the logical lifetime without perturbing the computational state.
However, many challenges remain to be addressed. In this work, $\chi_{02}$ is minimized for erasure detection via careful design and screening of device parameters, but extending to larger scales requires a more robust design that allows \textit{in situ} tuning of $\chi_{02}$ at $\Phi_\text{ext} = 0$. The large error rates of the erasure detection clearly call for improvements in measurement fidelity. Most importantly, we are unable to reach a regime where erasure errors dominate over Pauli errors due to the rapid dephasing of the logical qubit. This highlights the need for improved fabrication or the application of spin-locking techniques~\cite{Kubica2023}. Nonetheless, the approach is naturally scalable, as it can be readily adapted to implement error-preserving entangling gates using the FTF architecture~\cite{Ding2023, JHWang2025}, offering a hardware-efficient path toward erasure-based error correction.

\begin{acknowledgments}

The authors acknowledge Feng Wu and Huihai Zhao for insightful discussions. The Josephson parametric amplifier used in this work was provided by Beijing Academy of Quantum Information Sciences. X. Ma also acknowledges Yun Fan, Jun You, and the broader Heifei Zhileng Cryogenic Technology Co., Ltd. for their assistance in establishing the experimental infrastructure. This research is supported by the Guangdong Provincial Quantum Science Strategic Initiative (Grant No. GDZX2407001). H.-F. Yu acknowledges support from the National Natural Science Foundation of China (Grants No. 92365206) and the Innovation Program for Quantum Science and Technology (Grants No. 2021ZD0301802).

\end{acknowledgments}

\bibliographystyle{naturemag} 
\bibliography{ref}

\end{document}


\title{Supplementary Material for \\
``Converting qubit relaxation into erasures with a single fluxonium''}
\maketitle

\bookmarksetup{startatroot}

\section{Logical single-qubit gates}

To fully leverage the noise bias~\cite{Wu2022,Kubica2023} provided by erasure conversion, it is essential to implement logical gates that preserve this bias throughout the computation. For our device, this requires performing logical single-qubit gates at the zero-flux position, so that decay errors occurring during the gate sequence, as well as leakage to the $\ket{1}$ state, are in principle detectable using the erasure-conversion technique discussed in the main text. However, this is complicated by the fact that $\ket{0}\leftrightarrow \ket{2}$ is a forbidden transition at $\Phi_\text{ext} = 0$. In this work, we develop two approaches to work around this limitation. 

The first approach relies on a sequential rotation of the qubit through the allowed single-photon transitions. Specifically, we can construct an arbitrary rotation in the $\{\ket{0}, \ket{2}\}$ manifold via $\mathcal{R}_\theta^{02}(\varphi) = \mathcal{R}_\pi^{01}(0) \mathcal{R}_\theta^{12}(\varphi) \mathcal{R}_\pi ^{01}(0)$, where $\mathcal{R}_\theta^{ij}(\varphi)$ denotes a rotation in the $\{\ket{i}, \ket{j}\}$ subspace of angle $\theta$ and phase $\varphi$. Being relatively straightforward to calibrate, this gate sequence is used in the state preparations of experiments discussed in Fig.1 through Fig.3 of the main text. However, due to its construction, the qubit spends considerable time in the erasure state $\ket{1}$. During this time, photons in the resonator, either from active erasure detection or residual thermal excitation, can rapidly dephase the qubit in the $\{\ket{1},\ket{2}\}$ subspace, which in turn leads to dephasing of the logical qubit. Consequently, such gate schemes are ill-suited for more advanced characterization of the logical decoherence rate, such as spin-echo experiments.

\begin{figure}
    \centering
    \includegraphics{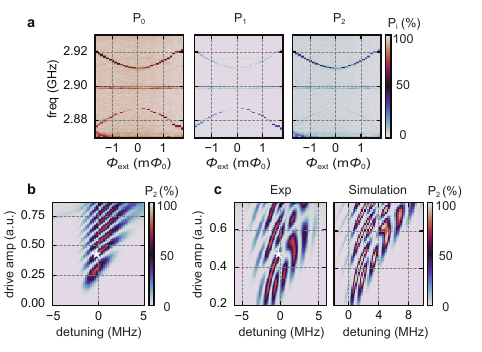}
    \caption{
    \textbf{Logical single-qubit gates}. 
    \textbf{a,} Spectroscopic measurement of the two-photon transitions. 
    Different panels display the probability $P_i$ of finding the qubit in the final $\ket{i}$, respectively. Among the spectroscopic features that tunes with $\Phi_\text{ext}$, the top one corresponds to the $\ket{0} \leftrightarrow \ket{2}$ transition, and the bottom one corresponds to the $\ket{0} \leftrightarrow \ket{1}$ transition. 
    \textbf{b,} At $\Phi_\text{ext} = 0$, the Raman chevron of a $480~$ns pulse is measured as a function of both drive amplitude and the frequency detuning from the $\ket{0}\leftrightarrow \ket{2}$ transition. Consequently, at the position indicated by the white star, we find a logical $\pi$-pulse for the computational manifold.
    \textbf{c,} Similarly, we calibrate a logical $\pi/2$-pulse (white star) by measuring (left panel) the Raman chevrons from two identical, $480~$ns pulses applied consecutively. The origin of the complex structures is unclear, but a numerical simulation (right panel) of the driven dynamics reproduces similar patterns.
    }
    \label{fig:s1}
\end{figure}

Instead, for the experiments described in Fig. 4 of the main text, we implement logical single-qubit rotations via a two-photon Raman transition.
\autoref{fig:s1}\textbf{a} depicts the spectroscopic measurement of these two-photon transitions as a function of the qubit's external flux. The features appear at half the frequency of the corresponding transitions because they are excited using a single, strongly driven microwave signal. Logical end-of-line measurements allow us to clearly identify the addressed qubit transition. At $\Phi_\text{ext} = 0$, we start from $\ket{0}$ and calibrate a logical $\pi$-pulse by measuring the probability of finding the qubit in $\ket{2}$ after driving it for $480~$ns with a microwave pulse of varying frequency and amplitude (\autoref{fig:s1}\textbf{b}). The tilt of the Raman chevrons toward higher drive amplitudes at increased pump frequencies arise from the ac-Stark shift imparted by the strong pump. Similarly, we calibrate a $\pi/2$-pulse by applying two consecutive drives of $480~$ns before measuring the probability of finding $\ket{2}$. 
\autoref{fig:s1}\textbf{c} shows the resulting Raman chevons. While the origin of the complex structure is unclear, numerical simulations of the driven dynamics reproduce similar patterns. In this work, the logical single-qubit gates performed using the Raman transition is relatively slow at $480~$ns. Going forward, better filtering and attenuation on the drive line could increase gate speed. Alternatively, we envision that driving the Raman transition with a combination of pump tones close to the single-photon transitions could also reduce the gate time.

\section{QNDness of erasure detections}

Because erasure qubit operations rely on repeated erasure detections, these measurements must exhibit a high degree of quantum non-demolition (QND) character. In particular, the measurement must leave the qubit undisturbed when it resides within the logical subspace. By contrast, non-QND effects on the erasure state is less detrimental, since a qubit flagged as erased will be reinitialized into the computational basis~\cite{Gu2025}. In our experiment, we therefore require the erasure detection to exhibit a high degree of QNDness if the qubit is in the $\{\ket{0},\ket{2}\}$ manifold.

\begin{figure}
    \centering
    \includegraphics{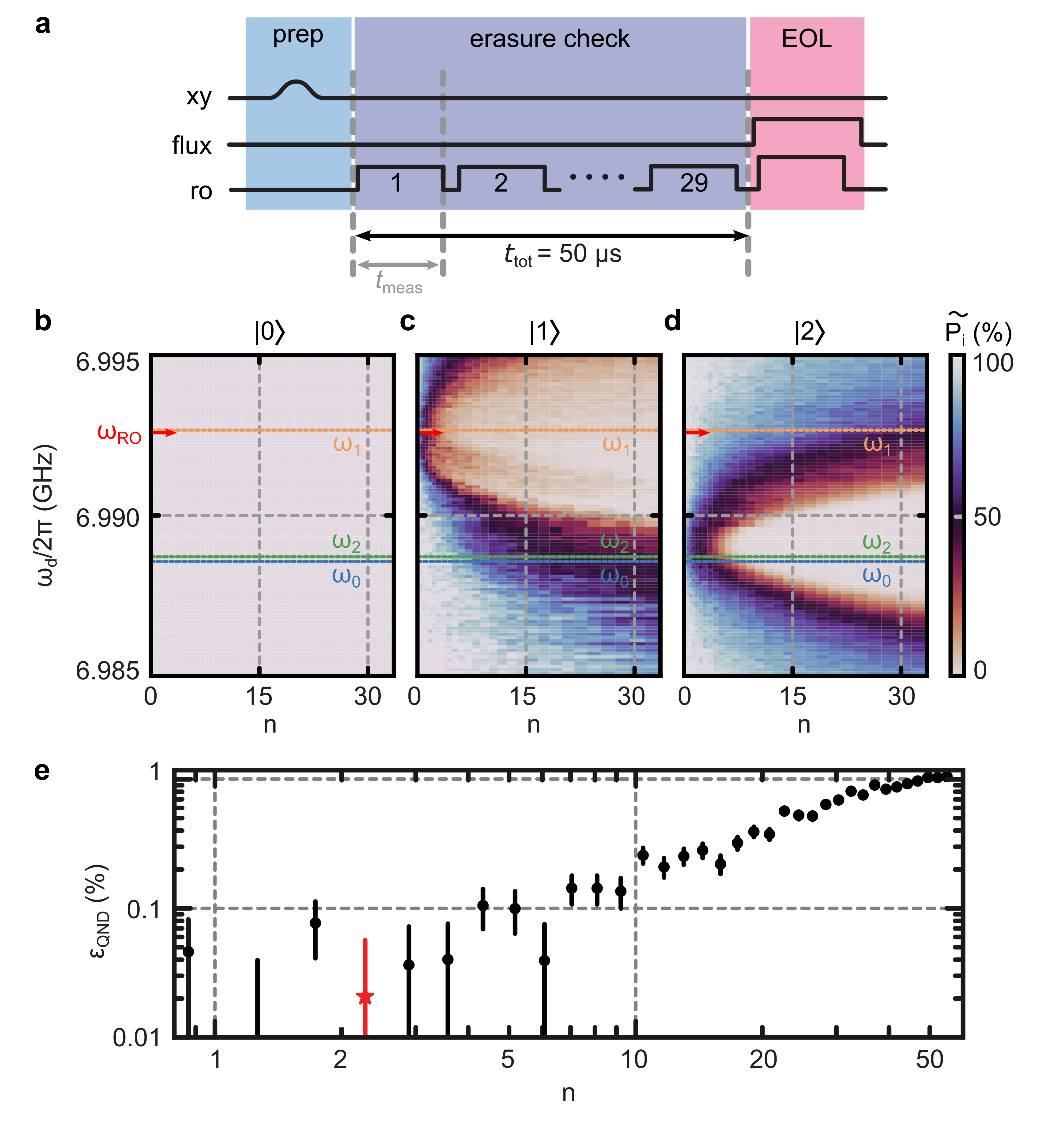}
    \caption{
    \textbf{QNDness of erasure detections}. 
    \textbf{a,} To characterize the QNDness of the erasure detection, the qubit initialized in state $\ket{i}$ is subjected to a sequence of 29 erasure detections for a total duration of $t_\text{tot} = 50~\mu$s before a final EOL measurement finds the probability $P_{\braket{i|i}}$ that the qubit remains in its initial state. 
    \textbf{b,} Starting with $\ket{0}$, the probability $P_{\braket{0|0}}$ is measured as a function of readout strength $n$ and frequency $\omega_d$. We normalize this probability to the case where the erasure detection is off ($n=0$), and plot the resulting $\tilde{P_0}(n,\omega_d) = P_{\braket{0|0}}(n,\omega_d)/P_{\braket{0|0}}(n=0)$. \textbf{c,d,} Similarly, we find the normalized probability $\tilde{P_1}$ (\textbf{c}) and $\tilde{P_2}$ (\textbf{d}) when the qubit is initilized in $\ket{1}$ and $\ket{2}$ respectively. The dotted horizontal lines indicate the qubit-dressed resonator frequency $\omega_i$, and the red arrows indicate the erasure detection frequency $\omega_\text{RO} = 6.993~$GHz used in the main text.
    \textbf{e,} At $\omega_\text{RO}$, we find the non-QND error $\varepsilon_\text{QND}$ per erasure check as a function of measurement strength. For the measurement strength used in the main text (red), $\varepsilon_\text{QND} < 0.1\%$ per check.
    }
    \label{fig:s2}
\end{figure}

To this end, we characterize the QNDness of our erasure measurement as a function of the readout frequency and readout power, as shown in \autoref{fig:s2}\textbf{a}. After initializing the qubit in one of its three lowest energy eigenstates $\ket{i}$ ($i \in [0,2]$), we perform a sequence of 29 erasure detections with varying drive frequency $\omega_d$ and measurement strength, expressed as the intra-cavity photon number $n$ if the drive were on resonance (see \autoref{sec:measurement-induced dephasing}). After a total delay time of $t_\text{tot} = 50~\mu$s, with each erasure detection lasting for $t_\text{meas} = 1.6~\mu$s, we perform an EOL measurement and extract the probability to find the qubit in its initial state, $P_{\braket{i|i}}(n,\omega_d)$. To separate measurement-induced non-QND effects from SPAM errors and the natural qubit decay during $t_\text{tot}$, we normalize this probability to its value at $n = 0$, where no measurement is applied and thus no non-QND effects can occur. We plot this normalized probability, $\tilde{P_i}(n,\omega_d) = P_{\braket{i|i}}(n,\omega_d)/P_{\braket{i|i}}(n=0)$ in \autoref{fig:s2}\textbf{b -- d}, for qubit initialized in $\ket{0}$, $\ket{1}$, and $\ket{2}$ respectively. 
For a given measurement strength $n$, the non-QND effect reaches its maximum for the initial state $\ket{i}$ when the erasure detection is performed on resonance with the $\ket{i}$-state dressed resonator frequency ($\omega_i$, indicated dashed horizontal lines). Away from this resonance, the filtering of the resonator reduces the actual intra-cavity photon number, thereby reducing the measurement backaction.

Taking advantage of this resonator filtering, we enhance the QNDness in the computational space by performing erasure detection at a frequency $\omega_\text{RO} = 2\pi \times 6.993~$GHz (horizontal red arrow) close to $\omega_1$. 
Because the dispersive shift $\chi_{01} = 2\pi \times -4.096~\text{MHz}$ is large compared to the cavity linewidth $\kappa = 2\pi \times 1.02~\text{MHz}$, this readout frequency is also well-suited for distinguishing $\ket{1}$ from the $\{ \ket{0},\ket{2} \}$ manifold~\cite{Gambetta2006}.  
\autoref{fig:s2}\textbf{e} shows the measurement-induced non-QND error in the computational space $\epsilon_\text{QND}$ per erasure check as a function of measurement strength $n$ at the chosen readout frequency, extracted according to 
\begin{equation}
    1-\epsilon_\text{QND} = \left(\frac{\tilde{P_0} + \tilde{P_2}}{2}\right)^{1/m}
\end{equation}
where $m = 29$ is the total number of erasure checks performed in the measurement sequence. For the measurement strength $n=2.3$ used in the main text (red star), we can therefore place an upper-bound on the non-QND error $\epsilon_\text{QND} < 0.1\%$ per check.

\section{Logical lifetime vs. $t_\text{EC}$}

\begin{figure}[h!]
    \centering
    \includegraphics{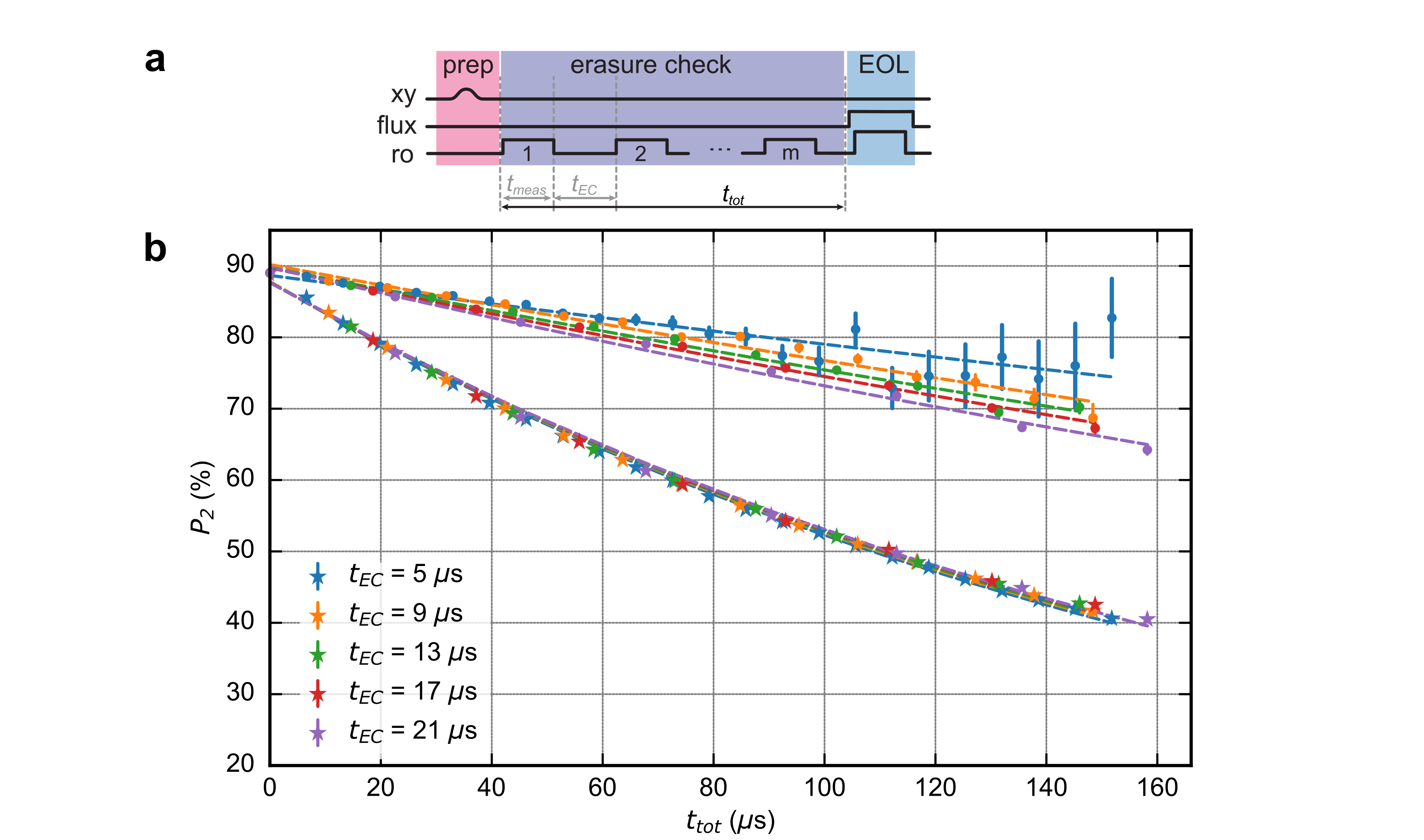}
    \caption{\textbf{Logical lifetime vs. $t_\text{EC}$}. 
    \textbf{a,} Measurement sequence reproduced from Fig.~3\textbf{e} of the main text. To characterise erasure conversion, a qubit initialized in $\ket{2}$ is subjected to a series of $m$ erasure detections, each lasting $t_\text{meas} = 1.6~\mu$s and separated by a variable interval $t_\text{EC}$, resulting in a total delay of $t_\text{tot}$, before an EOL measurement is performed.
    \textbf{b,} Post-selecting on the absense of erasure events, the logical lifetime of the qubit increases as $t_\text{EC}$ is reduced.
    }
    \label{fig:s3}
\end{figure}

To characterize the effect of $t_\text{EC}$ on the logical lifetime, we follow the protocol in Fig.~3\textbf{e} of the main text (reproduced in \autoref{fig:s3}\textbf{a}). The qubit is initialized in $\ket{2}$, subjected to $m$ erasure checks separated by a variable delay $t_\text{EC}$ over a total duration $t_\text{tot}$, and finally measured at the end-of-line to determine the probability of it remaining in $\ket{2}$ (\autoref{fig:s3}\textbf{b}).
The ratio of $t_\text{EC}$ to the characteristic lifetime of the erasure state, $T_{\ket{1}} = 1/(\Gamma_{12} + \Gamma_{10}) \approx 75.8~\mu$s, determines the likelihood that an erasure event remains in $\ket{1}$ at the next erasure detection. Shorter $t_\text{EC}$ reduces the chance that an erasure escapes detection by spontaneously returning to the computational basis. Performing multiple erasure checks while the qubit resides in $\ket{1}$ further lowers the probability of missing an erasure due to misassignment. Consequently, reducing $t_\text{EC}$ increases the logical lifetime obtained after post-selecting on the absence of erasure events. Note that the data for $t_\text{EC} = 5~\mu$s in \autoref{fig:s3}\textbf{b} is identical to the data reported in Fig.~3\textbf{g} of the main text.

\section{Measurement-induced dephasing}\label{sec:measurement-induced dephasing}

As described in the main text, we extract the pure-dephasing rate $\Gamma_\varphi = 1/T_{2\mathrm{L}} - 1/(2T_{1\mathrm{L}})$ (Fig.~4\textbf{b} of main text) as a function of the measurement frequency $\omega_d$ for a given measurement strength $\varepsilon_{\text{d}}$. To calibrate $\varepsilon_{\text{d}}$, we initialize the qubit in $\ket{0}$, and measure the ac-Stark shift $\delta\omega_{01}$ induced by a resonant probe of varying power on the $\ket{0}$-dressed resonator. This allows us to express the applied measurement strength in terms of the corresponding on-resonance intra-cavity photon number $n$.

Under an actual measurement pulse, the actual photon number $n_i$ inside the resonator depends on the qubit state $\ket{i}$ via the qubit-dressed resonator frequency $\omega_i$,
\begin{equation}
n_{i} =  
    \frac{\varepsilon_{\text{d}}^{2}}{\kappa^{2}/4 + (\omega_i - \omega_d)^2}
     \equiv 
     \frac{n\kappa^2}{\kappa^2 + 4(\omega_i- \omega_d)^2},
\end{equation}
where $\kappa$ is the cavity linewidth. These photon numbers are related to the measurement-induced dephasing according to~\cite{Gambetta2006},
\begin{equation}
    \Gamma_{\text{m}} = \frac{(n_{0} + n_{2})\,\kappa\,\chi_{02}^{2}}
   {\kappa^{2} + \chi_{02}^{2} + 4(\omega_r^0 - \omega_d)^{2}},
\end{equation}
where $\omega_r^0$ is the bare resonator frequency. 

Using the above relations, we fit the experimentally extracted pure-dephasing rates $\Gamma_\varphi = \Gamma_\text{m} + \Gamma_\varphi'$ at each measurement strength $n$. The only free parameters in the fits are the residual dephasing rate $\Gamma_\varphi'$ arising from non-measurement sources and the bare resonator frequency $\omega_r^{0}$. The resulting fits are shown as solid lines in Fig.~4\textbf{b} of the main text. Based on the fit result, we extract a dephasing rate $\Gamma_m \approx 2\pi\times 45$~Hz for our erasure detection of strength $n = 2.3$ performed at $\omega_\text{RO} = 2\pi\times 6.993~$GHz. Therefore, given our erasure detection time of $t_\text{meas} = 1.6~\mu$s, the measurement-induced dephasing error is $\epsilon_{\varphi} = 1-\text{exp}(-\Gamma_\text{m} t_\text{meas}) \approx 7.2\times 10^{-5}$ per check.

\section{Adiabatic limit for flux pulses}

In this work, we use flux pulses to perform EOL measurements away from the zero-flux position. To ensure that the qubit population remains unchanged during this flux excursion, the ramp must remain adiabatic with respect to the qubit dynamics. In particular, because $\omega_{12} \ll \omega_{01}$, the $\ket{1} \leftrightarrow \ket{2}$ transition is the most vulnerable to non-adiabatic excitations.

We estimate the corresponding non-adiabatic error using the Landau–Zener model. Working in the basis of the persistent-current states that hybridize to form the fluxonium~\cite{Ma2024}, we write a simplified Hamiltonian for the flux-ramped two-level system in the ${\ket{1},\ket{2}}$ manifold as
\begin{equation}
    H/\hbar = \omega_{12} \sigma_x + \dot{\omega}_{12} t \sigma_z,
\end{equation}
where $\dot{\omega}_{12}$ is the linear ramp rate of the qubit frequency as the flux is swept from large negative to large positive detuning. The resulting probability of a non-adiabatic transition is
\begin{equation}
    \epsilon_{na} = \text{exp}\left( -\frac{\pi \omega_{12}^2}{\dot{\omega}_{12}}  \right).
\end{equation}

In this work, we perform EOL measurements at $\Phi_\mathrm{ext} = 1.3~\mathrm{m}\Phi_0$, which, based on our measured qubit parameters, corresponds to a maximum qubit-frequency shift of $\delta \omega_{12} \approx 2\pi \times 8.6~\text{MHz}$. For our $10~\text{ns}$ ramp duration, this yields a non-adiabatic error of only $\epsilon_{na} \sim 10^{-25}$, confirming that the flux excursion is safe within the adiabatic regime.

\bibliographystyle{naturemag} 
\bibliography{ref}